\documentstyle{article}

\if@twoside
\else
\fi
\advance\textheight by \topskip

\makeatletter
\def\@oddhead{\hfill\verb+aipproc2.tex+}
\let\@evenhead\@oddhead
\def\se{\vskip3pt plus1pt minus1pt\setbox0=\hbox to\hsize\bgroup\hss
        \vrule width.5pt
        \vbox\bgroup \hrule width \hsize height.5pt
        \vskip3pt\hbox to\hsize\bgroup\hss\vbox\bgroup\advance\hsize by-9pt
        \columnwidth\hsize\small}
\def\ee{\par\egroup\hss\egroup\vskip3pt\hrule width\hsize height.5pt\egroup
        \vrule width.5pt\hss\egroup
        \box0 \vskip3pt plus1pt minus1pt}
\def\latexe{\LaTeX\kern.15em 2${}_{\textstyle\varepsilon}$}

\makeatother

\begin{document}
\sloppy		

\begin{titlepage}
\vspace*{0pt plus1fill}
\begin{center}
\LARGE\bf
Minimal Model for Neutrino Masses and Mixings\\[1pc]
Paul H. Frampton
\end{center}
\vspace*{0pt plus1fill}
\vspace*{0pt plus1fill}
\hbox to\hsize{\hfil
\vbox{\offinterlineskip\hrule height 2pt\vskip4pt
\Large\sf\setbox0=\hbox{May 4, 2000. San Juan, Puerto Rico.}
\hbox to\wd0{\hfil Talk at Second Tropical Conference 
\hfil}
\hbox to\wd0{\hfil on Particle Physics and Cosmology,
\hfil}
\vskip4pt
\box0
\vskip4pt \hrule height 2pt}%
\hfil}
\end{titlepage}

\newpage

\begin{center}
\LARGE\bf
Abstract
\vspace{2pc}
\end{center}
Working in the framework of three chiral neutrinos
with Majorana masses, we investigate a scenario first
realized in an explicit model by Zee: that the
neutrino mass matrix is strictly
off-diagonal in the flavor basis,  with all
its diagonal entries precisely zero.
This CP-conserving ansatz leads to two relations among the
three mixing angles
$(\theta_1, \theta_2, \theta_3)$ and
two squared mass differences.
We impose the constraint $|m_3^2 - m_2^2| \gg |m^2_2 - m_1^2|$
to conform with experiment, which requires the $\theta_i$
to lie nearby one of four 1-parameter domains in $\theta$-space. We
exhibit the implications for solar and
atmospheric neutrino oscillations in  each of these cases.
A unique version of the Zee {\it ansatz\/} survives
confrontation with  experimental
data, one which necessarily involves
 maximal just-so vacuum oscillations
of solar neutrinos.

\newpage

\def\contentsname{\LARGE\bf Contents}

{\large
\tableofcontents
}

\twocolumn

\section{Introduction}

In this talk I will describe a paper published last year\cite{FG}
with Shelly Glashow; the work was done at the 
Korean Institute for Advanced
Study, Seoul, in June 1999 and the topic is what is the simplest
way to extend the minimal standard model, where neutrinos are
by definition massless, to accommodate the compelling evidence
especially from the SuperKamiokande experiment
for non-zero neutrino mass.

\section{The Situation of the Data}

The minimal standard model involves three chiral neutrino states, but
it does not admit renormalizable interactions that
can generate neutrino masses. Nevertheless, experimental
evidence suggests that
both solar and atmospheric neutrinos display flavor oscillations,
and  hence that neutrinos do have mass.
Two very different neutrino squared-mass differences are required to fit
the data:
\begin{equation}
10^{-11}{\rm eV^2} \le  \Delta_s \le 10^{-5}{\rm eV^2}
~~~ {\rm and} ~~~ \Delta_a\simeq 10^{-3}{\rm eV^2}
\label{edata}
\end{equation}
where the neutrino masses $m_i$ are ordered  such that:
\[
\Delta_s \equiv | m_2^2-m_1^2 | {\rm and} \Delta_a \equiv \vert m_3^2-m_2^2
\vert \simeq \vert m_3^2-m_1^2 \vert
\]
and the subscripts $s$ and $a$ pertain to solar and atmospheric
oscillations respectively.
 The large
uncertainty in $\Delta_s$ reflects  the several potential  explanations
of the observed solar neutrino flux: in terms of vacuum oscillations or
large-angle  or  small-angle MSW solutions,  but in
every case the two independent squared-mass differences must be widely
spaced with
\[
r\equiv \Delta_s/\Delta_a < 10^{-2}
\]

In a three-family scenario, four neutrino
mixing parameters suffice to describe neutrino oscillations,
akin to the four
Kobayashi-Maskawa parameters in the quark sector.\footnote{Two
additional convention-independent
phases are measurable in principle, but they ordinarily do not affect
neutrino oscillations\cite{J,G}}

Solar neutrinos
may exhibit an energy-independent time-averaged suppression due to
$ \Delta_a $, as well as energy-dependent oscillations depending on
$\Delta_s/E$. Atmospheric neutrinos may exhibit oscillations due to
$\Delta_a$, but they
are almost entirely unaffected by $\Delta_s$. It is convenient
to define
neutrino mixing angles as follows:
\[
\left( \begin{array}{c}\nu_e \\\nu_{\mu} \\ \nu_{\tau} \end{array} \right)
= X 
\left( \begin{array}{c} \nu_1 \\ \nu_2 \\ \nu_3 \end{array} \right)
\]
where
\[ X =
\left( \begin{array}{ccc}c_2c_3 & c_2s_3 & s_2e^{-i\delta} \\
+c_1s_3 +s_1s_2c_3e^{i\delta}& -c_1c_3-s_1s_2s_3e^{i\delta}&-s_1c_2 \\
+s_1s_3 -c_1s_2c_3e^{i\delta}& -s_1c_3-c_1s_2s_3e^{i\delta}&+c_1c_2
\end{array} \right) \]
with $s_i$ and $c_i$ standing for sines and cosines of $\theta_i$.
For neutrino masses satisfying Eq.(\ref{edata})
the vacuum survival  probability of solar neutrinos is\cite{GG}

\begin{equation}
P(\nu_e\rightarrow\nu_e)\big\vert_s \simeq 1-{\sin^2{2\theta_2}\over2}
- \cos^4{\theta_2}\sin^2{2\theta_3}sin^2{(\Delta_s R_s/4E)}
\label{eproba}
\end{equation}
whereas the transition probabilities  of atmospheric neutrinos are:

\begin{eqnarray}
P(\nu_\mu  \rightarrow\nu_\tau)\big\vert_a  & \simeq &
\sin^2{2\theta_1}\cos^4{\theta_2}\,\sin^2{(\Delta_a R_a/4E)} \nonumber  \\
 P(\nu_e \leftrightarrow\nu_\mu)\big\vert_a & \simeq &
\sin^2{2\theta_2}\sin^2{\theta_1}\,\sin^2{(\Delta_aR_a/4E)} \nonumber \\
 P(\nu_e \rightarrow\nu_\tau)\big\vert_a  & \simeq &
\sin^2{2\theta_2}\cos^2{\theta_1}\,\sin^2{(\Delta_aR_a/4E)} \nonumber \\
\label{atmos}
\end{eqnarray}

None of these probabilities depend on $\delta$, the measure of CP violation.

\section{Extending the Standard Model}

Let us turn to the origin of neutrino masses.
Among the many
renormalizable and gauge-invariant extensions of the standard model
that can do the trick are:

\begin{itemize}

\item The introduction of a complex triplet of mesons
($T^{++},\, T^+,\,T^0$) coupled bilinearly to pairs of lepton
doublets.
They must also couple bilinearly to the
Higgs doublet(s) so as to avoid spontaneous $B-L$ violation and the
appearance of a massless and experimentally excluded  majoron.
 This mechanism can generate an arbitrary  complex symmetric
Majorana mass matrix for neutrinos.

\item The introduction of singlet counterparts to the
neutrinos with very large Majorana masses. The interplay between these mass
terms and those generated by the Higgs boson---the so-called
see-saw mechanism---yields an arbitrary
but naturally small Majorana neutrino mass matrix.

\item The introduction of a charged singlet meson $f^+$
coupled antisymmetrically to pairs of lepton doublets, {\it and\/} a
doubly-charged singlet meson $g^{++}$ coupled bilinearly both to pairs of
lepton singlets and to pairs of $f$-mesons. An arbitrary Majorana neutrino
mass matrix is generated in two loops.

\item The introduction  of a charged singlet meson $f^+$
coupled antisymmetrically to pairs of lepton doublets {\it and\/} (also
antisymmetrically) to a pair of Higgs doublets.  This
simple mechanism was first proposed by Tony Zee\cite{zee}
and
results (at one loop) in a Majorana mass matrix
in the flavor basis ($e,\mu,\tau$)
 of a  special form:
\begin{equation}
{\cal M}= \left( \begin{array}{ccc} 0 & m_{e\mu} & m_{e\tau} \\
 m_{e\mu} & 0 & m_{\mu\tau} \\
                   m_{e\tau} & m_{\mu\tau} & 0 \end{array} \right)
\label{ansatz}
\end{equation}

\end{itemize}

\noindent
In extending the standard model one crucial requirement is
anomaly cancellation ({\it e.g.} \cite{CC}) and this is
satisfied by all
the above four scenarios.
We focus here on the last scenario. In particular,
we adopt the Zee {\it ansatz\/} for $\cal M$
without committing ourselves to the Zee mechanism for its origin.
Related  discussions of Eq.(\ref{ansatz})
appear elsewhere\cite{J,ST}
The present work is essentially a continuation of \cite{J}.

Because the diagonal entries of $\cal M$ are zero,  the amplitude for
no-neutrino double beta decay vanishes at lowest order
\cite{zee} and this process cannot proceed at an observable rate.
Furthermore, the
parameters $m_{e\mu},\, m_{e\tau}$ and
$m_{\mu\tau}$ may be taken real and non-negative
without loss of generality, whence  $\cal M$ becomes  real as well as
traceless and
symmetric.  With this convention, the analog to the
Kobayashi-Maskawa matrix becomes orthogonal:
$\cal M$ is explicitly
CP invariant and $\delta=0$. But as we have noted,
it is well known\cite{GG} that the mere existence of
a squared-mass hierarchy  virtually precludes any detectable
manifestation of
CP violation in the neutrino sector.

The sum of the neutrino masses (the eigenvalues of $\cal M$)
vanishes:
\begin{equation}
 m_1+m_2+m_3=0
\label{etrace}
\end{equation}
An important result  emerges when the squared-mass hierarchy
Eqs.(\ref{edata}) is taken into account along with Eq.(\ref{etrace}).
In the limit $r\rightarrow 0$, two of the squared masses must be equal.

\section{The Logical Possibilities}

There are two possibilities.
 In case A, we have
$m_1+m_2 = 0$ and $m_3= 0$. This case arises
iff at least one of
the three parameters in $\cal M$  vanishes.
In case B, we have $m_1= m_2$  and  $m_3= -2m_2>0$. This case
arises iff the three parameters in $\cal M$ are equal to one another.
Of course, $r$ is small but it does not vanish:  in neither case can
the above relations among neutrino masses be strictly satisfied.
But they must be nearly
satisfied. Consequently  we  may deduce
certain approximate but useful
restrictions on the permissable values of the
neutrino mixing angles $\theta_i$. Prior to examining these restrictions,
we note that Eqs.(\ref{etrace}) and (\ref{edata})
exclude the possibility that the three neutrinos are nearly degenerate in
mass.  If the Zee
{\it ansatz\/} is
even approximately realized in nature,
no neutrino mass can exceed a small fraction of
an elecron volt in magnitude and
neutrinos  are unlikely to contribute  significantly to
the dark matter of the universe.

We first consider case A. The relation $m_1+m_2=0$ may be obtained in three
ways depending on which parameter in $\cal M$ is set to zero.
 If $m_{e\mu}=0$, the quantum number $L_\tau-L_e-L_\mu$ is
conserved. It follows that  $\cos{\theta_1}=0$ and
$\theta_3=\pi/4$. We see from Eq.(\ref{atmos}) that atmospheric $\nu_\mu$'s
oscillate exclusively into $\nu_e$'s
and {\it vice versa.} This subcase (or any nearby assignment of mixing angles)
is strongly disfavored by
SuperKamiokande  data\cite{superK}.

Alternatively we may set $m_{e\tau}=0$ to obtain
conservation of $L_\mu-L_\tau-L_e$.
 For this subcase, we obtain $\sin{\theta_1}=0$ and
$\theta_3= \pi/4$. We see from Eq.(\ref{atmos})
that atmospheric $\nu_\mu$'s do
not oscillate at all. This subcase (or any nearby assignment of mixing angles)
is also strongly disfavored by
SuperKamiokande data\cite{superK}.

\bigskip

The last and best (\cite{J}) version of Case A has $m_{\mu\tau} = 0$
and leads to conservation of $L_e-L_\mu-L_\tau$.
 For this subcase, we obtain $\sin{\theta_2}=0$ and
$\theta_3=\pi/4$. We see from Eq.(\ref{eproba})  that solar
neutrino oscillations are maximal:
\begin{equation}
P(\nu_e\rightarrow\nu_e)\big\vert_s=
1-\sin^2{(\Delta_s R_s/4E)}
\label{esolarosc}
\end{equation}
Moreover, we see from Eq.(\ref{atmos})
that atmospheric $\nu_\mu$'s oscillate exclusively into $\nu_\tau$'s with
the unconstrained mixing angle $\theta_1$:
\begin{eqnarray}
P(\nu_\mu\rightarrow \nu_\tau)\big\vert_a
& = & \sin^2{2\theta_1}\,\sin^2{(\Delta_a
R_a/4E)} \nonumber \\
P(\nu_\mu \leftrightarrow \nu_e)\big\vert_a = 0 & ~~~ &
\quad\quad P(\nu_e\rightarrow \nu_\tau)\big\vert_a=0 \nonumber \\
\label{eatmososc}
\end{eqnarray}

This implementation of the Zee {\it ansatz\/} is compatible with experiment:
It predicts
maximal solar oscillations (without an energy-independent term)
and it is consistent with the just-so vacuum oscillation
hypothesis\cite{BWP}. However, it
is evidently not compatible with the small-angle MSW explanation of solar
neutrino data. Neither is it compatible with the large-angle MSW solution,
because it predicts virtually  maximal solar neutrino oscillations.
If $m_{\mu\tau}$ is permitted to depart slightly from zero so as to generate
a small finite value of $r\equiv\Delta_s/\Delta_a$,
the coefficient of the oscillatory term in Eq.(\ref{esolarosc}) will depart from
unity by a term of order $r^2\le  10^{-4}$. The resulting
solar-neutrino oscillations remain nearly maximal: they are
energy independent and
experimentally disfavored  unless $\Delta_s$ lies within the just-so domain.

We furthermore note that
the $m_{\mu\tau}\simeq 0$  version of the Zee {\it ansatz\/} admits
atmospheric neutrino oscillations
of the type $\nu_\mu\rightarrow \nu_\tau$ with any value of the
mixing angle  $\theta_1$.
At the same time, it  precludes  all oscillations involving
atmospheric $\nu_e$. These results are
 quite in accord with SuperKamiokande data.

Others have  noted\cite{barbieri}
 that the relation $m_1^2=m_2^2$ is preserved by radiative corrections for
case A. This is not necessarily true for case B,
 where   the relations satisfied by the neutrino masses,
 $m_3= -2m_2= -2m_1$, are not a consequence of a symmetry principle.
In any event, we argue  that case B cannot fit the data.
Ref.\cite{J} shows
that case B leads to the relation:
\begin{equation}
\tan^2{\theta_2}=1/2
\label{ejarl}
\end{equation}
Along with Eqs.(\ref{eproba}), this implies:
\begin{equation}
P(\nu_e\rightarrow \nu_e)\big\vert_a
=1-(8/9)\,\sin^2{(\Delta_aR_a/4E)}
\label{ecaseB}
\end{equation}

That is, we must have large (almost maximal)  oscillations of atmospheric
$\nu_e$. This result is strongly disfavored by
SuperKamiokande data\cite{superK}, so that
case B can be rejected without further ado.
\medskip

Our conclusion is simple.
We find that  one (and only one!) of the realizations of the Zee {\it
ansatz\/}  incorporating  a squared-mass hierarchy
is compatible with both solar and atmospheric neutrino data. It
corresponds to the assignments
$m_{e\mu}\simeq  m\cos{\theta_1}$, $m_{e\tau}\simeq
m\sin{\theta_1}$  and $ m_{\mu\tau}\ll m$, with $\theta_2\simeq 0$ and
$\theta_3\simeq \pi/4$.
Near this domain,
atmospheric electron neutrinos oscillate negligibly, while
atmospheric muon neutrinos oscillate into tau neutrinos with the arbitrary
mixing angle $\theta_1$. Solar neutrino oscillations are
very nearly maximal.
They can be described by vacuum oscillations,
but not by
MSW oscillations.  It is straightforward to
implement the Zee model
so as to conserve $L_e-L_\mu-L_\tau$
{\it exactly\/}  so as to obtain
$m_{\mu\tau}=0$ to all orders\cite{barbieri}.
Of course, this must not be done!
Some unspecified new physics
(beyond the introduction  of Zee's  $f^+$ meson) is required to lift the
degeneracy between $m_1^2$ and $m_2^2$ so as to yield an extreme hierarchy
of neutrino
squared-mass differences, with $\Delta_s\sim 10^{-8}\; \Delta_a$.
\bigskip\bigskip

\section{Acknowledgement}

\bigskip

This work was supported in
part by the U.S. Department of Energy
under grant number DE-FG02-97ER-41036.

\end{document}